# Materialized View Selection by Query Clustering in XML Data Warehouses

Hadj Mahboubi, Kamel Aouiche and Jérôme Darmont
ERIC – University of Lyon 2
5 avenue Pierre Mendès-France
69676 Bron Cedex, France
{hmahboubi,kaouiche,jdarmont}@eric.univ-lyon2.fr

## ABSTRACT

XML data warehouses form an interesting basis for decision-support applications that exploit complex data. However, native XML database management systems currently bear limited performances and it is necessary to design strategies to optimize them. In this paper, we propose an automatic strategy for the selection of XML materialized views that exploit a data mining technique, more precisely the clustering of the query workload. To validate our strategy, we implemented an XML warehouse modeled along the XCube specifications. We executed a workload of XQuery decision-support queries on this warehouse, with and without using our strategy. Our experimental results demonstrate its efficiency, even when queries are complex.

## 1. Introduction

Decision support applications nowadays exploit heterogeneous data from various sources. Furthermore, the development of the Web and the proliferation of multimedia documents contributed to the analysis of so-called complex data [8]. For instance, analyzing medical data may lead to exploit jointly information under various forms: patient records (classical database), medical history (text), radiographies, echographies (multimedia documents), physician diagnoses (texts or audio recordings), etc.

In this context, we have used XML in the process of integrating and warehousing complex data for analysis [7]. However, decision-support queries are generally complex because they involve several join and aggregation operations. In addition, native XML database management systems (DBMSs) present poor performances when the volume of data is very large and the queries are complex. Thus, it is crucial to design XML data warehouses that guarantee the best performance when accessing data. Indexing and view materialization are the most frequently used optimization techniques for this sake [12].

Materialized views are physical structures that improve data access time by precomputing intermediary query results. Then, end-user queries can be processed efficiently from the data stored within these views and do not need access the original data any more. Nevertheless, the use of materialized views requires additional storage space and induces some refreshing process overhead. So it is crucial to select only pertinent views.

In the context of relational data warehouses, several studies have been proposed to resolve the materialized view selection problem [1, 3, 4, 10, 11, 13, 18, 22, 23, 24, 25, 27]. The views that are relevant to materialize are selected to minimize the processing time of a given workload. This optimization is achieved under maintenance cost or storage space constraints [16]. The existing studies differ in several points:

1. the way of determining candidate views;
2. the framework used to capture relationships between candidate views;
3. the use of mathematical cost models *vs.* calls to the query optimizer;

4. the selection of views in a relational or multidimensional context;
5. multiple or simple query optimization;
6. theoretical or technical solutions.

The most recent approaches are workload-driven. They syntactically analyze the workload to enumerate the relevant candidate views [1]. By calling the query optimizer, they greedily build a configuration of the most pertinent views. A materialized view selection based on clustering has also been proposed [2]. This proposal exploits query clustering to determine a set of candidate views and cost models to choose pertinent views to materialize.

To the best of our knowledge, no such view materialization approach exists in XML databases and XML data warehouses in particular. Hence, we propose in this paper an adaptation of the query clustering-based relational view selection approach [2] to the XML context. Our approach clusters XQuery queries (instead of SQL queries) and builds candidate XML views that can resolve multiple similar queries belonging to the same cluster. New XML-specific cost models are used to define the XML views that are pertinent to materialize. To validate our proposal, we implemented an XML data warehouse in a native XML DBMS. It is indeed interesting to check whether native XML DBMSs could someday be able to compete with XML-compatible, relational DBMSs. Then, we measured the execution time of a decision-support query workload with and without using our strategy. Our experimental results show that the use of our strategy greatly improves query performance.

The remainder of this paper is organized as follows. We first present the context of this study in Section 2. Then we detail our materialized view selection strategy in Section 3. In order to validate our strategy, we present some experimental results in Section 4. Finally, we conclude and outline some research perspectives in Section 5.

# 2. Study context

## 2.1 XML data warehouse specification

Several studies have been proposed for designing and building XML data warehouses. For instance, Pokorny modeled a star schema in XML by defining dimension hierarchies as a set of logically connected collections of XML data, and facts as XML data elements [20, 21].

Park *et al.* also proposed an XML multidimensional model in which each fact is described by a single XML document and dimension data are grouped into a repository of XML documents [19].

Finally, Hummer *et al.* designed *XCube,* a family of templates allowing the description of a multidimensional structure, dimension and fact data for integrating several data warehouses into a virtual or federated data warehouse [14]. The federated templates are not directly related to XML warehousing, but they can be used to represent XML star schemas. XCube is organized as a set of modules or formats: *XCube Schema*, *XCube Dimensions* and *XCube Facts*, which respectively formalize the schema, the dimensions and the facts according to a star schema.

These studies use XML documents to manage or represent the facts and dimensions of an XML data warehouse. They actually help logically modeling a data warehouse. This allows the native storage of documents and their easy interrogation based on XML languages.

In this paper, we selected the XCube specification to model a reference XML data warehouse and apply our strategy. Indeed, in XCube, the authors proposed a simple structure for representing facts and dimensions in a star schema. They use one XML document to represent dimensions and one XML document to represent facts. In addition, they use another XML document representing warehouse metadata. The other proposals do not use XML documents for representing warehouse schema. However, we need these metadata to compute our cost models.

Thus, our data warehouse is composed of the following XML documents:

- ***Schema.xml*** specifies the data warehouse metadata;
- ***Dimensions.xml*** defines all the dimensions characterized by their attributes and values;
- ***Facts.xml*** specifies the facts, i.e., the identifiers of dimensions and the description of measures.

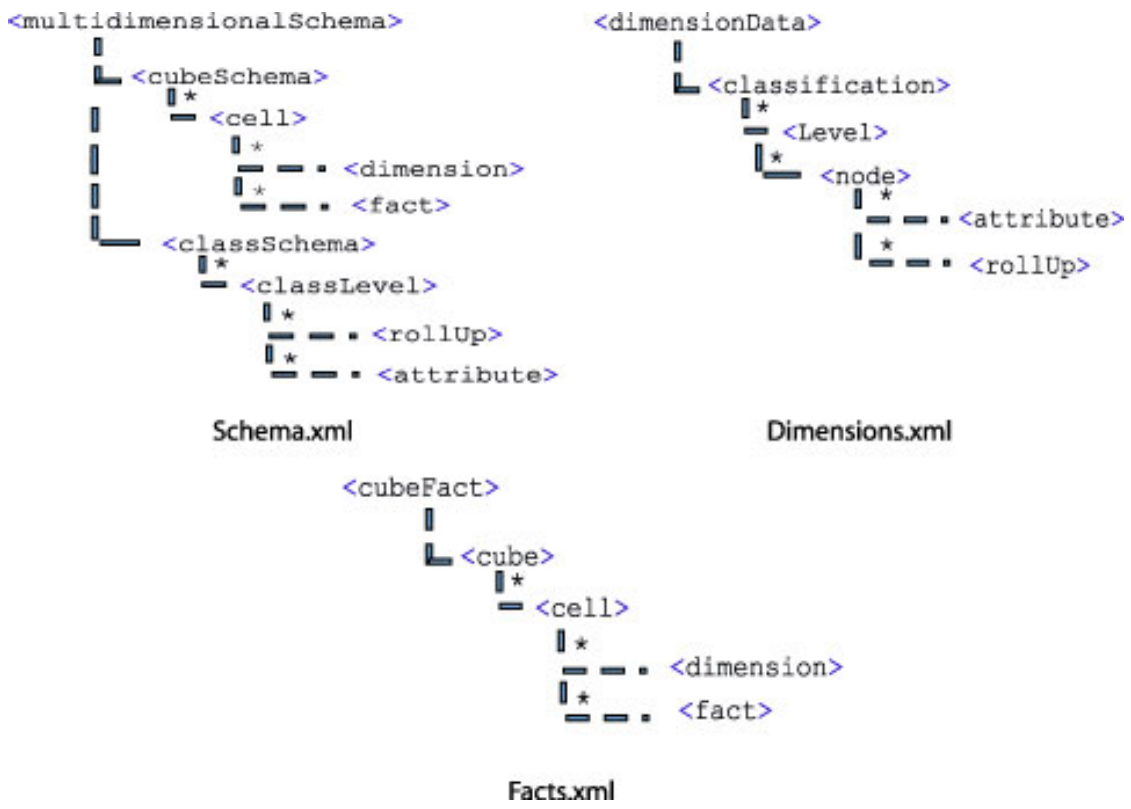

**Figure 1. XCube warehouse specification**

### 2.2 XML data warehouse interrogation

We selected the XQuery language [5] to formulate our decision-support queries because, unlike simpler languages such as XPath, it allows complex queries, including join queries over multiple XML documents, to be expressed with the *FLWOR* syntax. However, in our implementation, we had to extend *FLWOR* expressions with explicit *Group by* clauses to be able to formulate the decision-support queries we needed. Thus, we added the functions *Group by (attribute list)* and *Aggregation (aggregation operations, measure list)* to the XQuery syntax. Figure 2 provides an example of decision-support query with a multiple Group by clause.

```
for $a in document(Dimensions.xml)//dimensionData/classification/
                                     Level[@node='CUSTOMERS']/node,
    $x in document(Facts.xml)//CubeFacts/cube/Cell
where $a/attribute/@name='CUST_CITY'
        and $a/attribute/@value='Lyon'
        and $x/dimension/@node=$a/@id
        and $x/dimension/@id='CUSTOMERS'
Group by(@CUST_LAST_NAME,@CUST_POSTAL_CODE)
return sum (quantity)
```

**Figure 2. Decision-support XQuery example**

## 3. XML materialized view selection strategy

The architecture of our materialized view selection strategy is depicted in Figure 3. We assume that we have a workload composed of representative queries for which we want to select a configuration of materialized views in order to reduce their execution time. The first step is to build, from the workload, a clustering context. Then we define similarity and dissimilarity measures that help clustering together similar queries.

For each cluster, we build a set of candidate views. The last step exploits cost models that evaluate the cost of accessing data using views and the cost of their storage to build a final materialized view configuration.

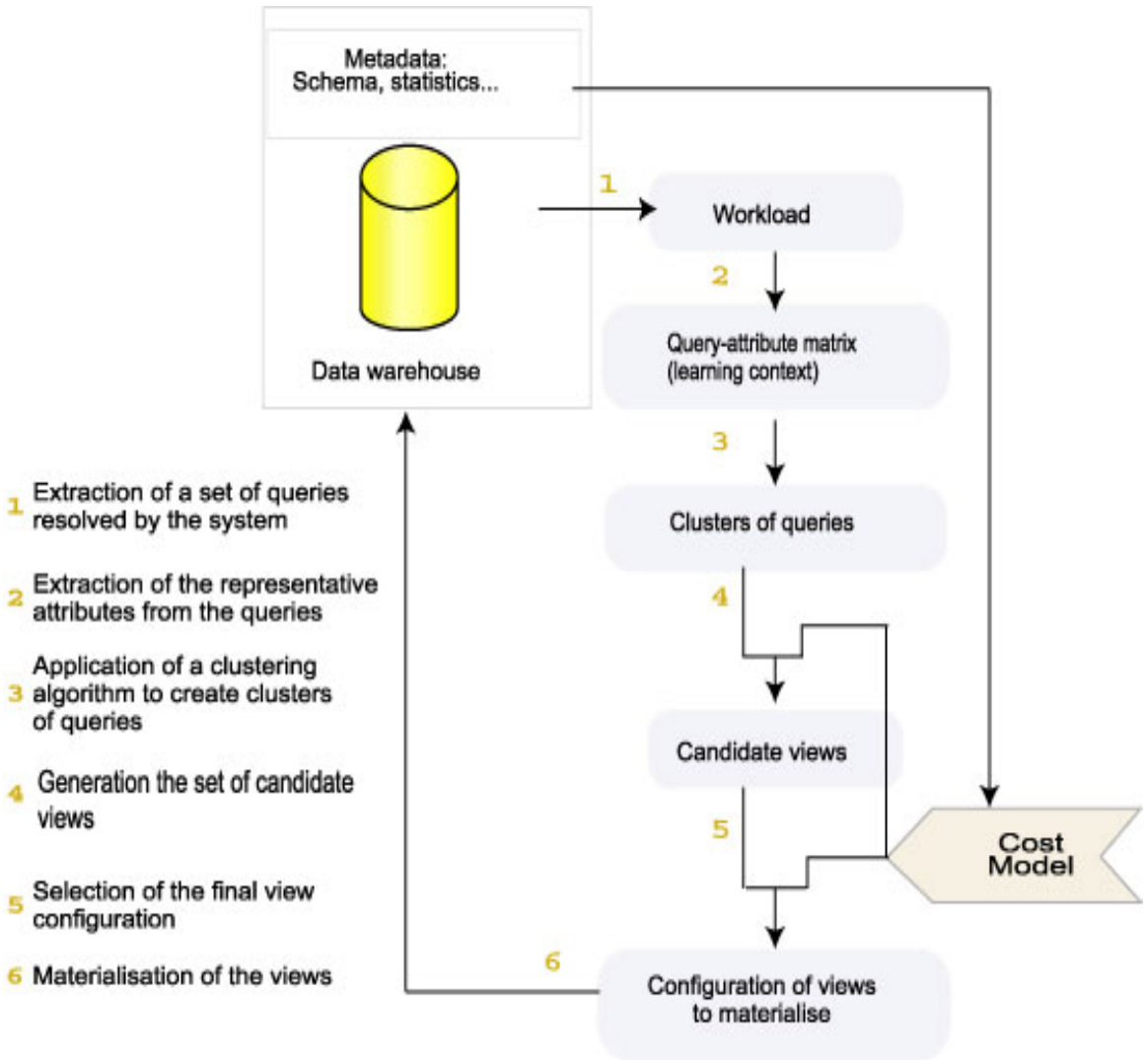

**Figure 3. Materialized view selection strategy**

```
q1  for $a in //dimensionData/classification/Level
    [@node='channels']/node,
    $b in //dimensionData/classification/Level
    [@node='customers']/node,
    $x in //CubeFacts/cube/Cell
    let $q := $b/attribute[@name='cust_first_name'],
    $s := $a/attribute[@name='channel_class']
    where $a/attribute[@name='channel_desc',@value='Internet']
    and $b/attribute[@name='cust_city',@value='Montpellier']
    and $x/dimension /@node=$a/@id
    and $x/dimension /@node=$b/@id
    and $x/dimension/@id='customers'
    and $x/dimension/@id='channels'
    group by(@cust_first_name,@channel_class)
    return @cust_first_name, @channel_class, sum(quantity)

q2  for $a in //dimensionData/classification/Level
    [@node='channels']/node,
    $b in //dimensionData/classification/Level
    [@node='customers']/node,
    $x in //CubeFacts/cube/Cell
    let $q := $b/attribute[@name='cust_first_name'],
    $s := $a/attribute[@name='channel_class']
    where $a/attribute[@name='channel_desc',@value= 'Internet']
    and $b/attribute[@name='cust_city',@value='Lyon']
    and $x/dimension /@node=$a/@id
    and $x/dimension/@node=$b/@id
    and $x/dimension/@id='customers'
    and $x/dimension/@id='channels'
    group by(@cust_first_name,@channel_class)
    return @cust_first_name,@channel_class, sum(quantity)

...
```

**Figure 4. Workload snapshot**

### 3.1 Query workload analysis

The workload that we consider is a set of *selection*, *join* and *aggregation* queries. Figure 4 gives a snapshot of this workload. The first step consists in extracting from the workload the representative attributes for each query. We mean by representative attributes those are present in *Where* (selection predicate attributes) and *Group by* clauses.

We store the relationships between the query workload and the extracted attributes in a "*query-attribute*" matrix. The matrix lines are the queries and the columns are the extracted attributes. A query $q_i$ is then seen as a line in the matrix that is composed of cells corresponding to representative attributes. The general term $q_{ij}$ of this matrix is set to one if extracted attribute $a_i$ is present in query $q_i$, and to zero otherwise. This matrix represents our clustering context. Table 1 shows the query-attribute matrix that is built from the workload snapshot from Figure 4.

### 3.2 Building the candidate view configuration

In practice, it is hard to search all the syntactically relevant views (candidate views) because the search space is very large [1]. To reduce the size of this space, we propose to cluster the queries. Hence, we group in a same cluster all the queries that are similar. Similar queries are the one having a close binary representation in the query-attribute matrix. Two similar queries can be resolved by using only one materialized view. We define similarity and dissimilarity measures that ensure that queries within a same cluster are strongly related to each others whereas queries from different clusters are significantly different.

#### 3.2.1 Similarity and dissimilarity measures

A query is described by the attributes extracted in the query analysis phase. We thus describe a query $q_i$ by a vector $q_i = \{q_{1i}, q_{2i}, ..., q_{pi}\}$, where $p$ is the number of attributes in the matrix. This description allows query comparison.

We define similarity (respectively, dissimilarity) between two queries $q_i$ and $q_j$ regarding attribute $a_k (k = \overline{1..p})$ in Formula 1 (respectively, Formula 2).

$$\delta_{sim}(q_{ki}, q_{kj}) = \begin{cases} 1 \, if \; q_{ki} = q_{kj} = 1 \\ 0 \; otherwise \end{cases} \qquad (1)$$

$$\delta_{dissim}(q_{ki}, q_{kj}) = \begin{cases} 1 \, if \; q_{ki} = q_{kj} \\ 0 \, if \; q_{ki} \neq q_{kj} \end{cases} \qquad (2)$$

Two queries $q_i$ and $q_j$ are similar regarding attribute $a_k$ if and only if $q_{ki} = q_{kj} = 1$, *i.e.*, $a_k$ is present in both queries. They are dissimilar if and only if $q_{ki} \neq q_{kj}$, *i.e.*, one of the two queries does not contain attribute $a_k$.

These measures can be extended to a set $A$ composed of $p$ attributes such that we get the degree of global similarity and dissimilarity between two queries. We thus define the similarity (respectively, dissimilarity) between two queries $q_i$ and $q_j$ according to all the matrix attributes $a_k$ in Formula 3 (respectively, Formula 4).

$$sim(q_i, q_j) = \sum_{j=1}^{p} \delta_{sim}(q_{ki}, q_{lj}) \quad (3)$$

$$0 \le sim(q_i, q_j) \le p$$

$$dissim(q_i, q_j) = \sum_{j=1}^{p} \delta_{dissim}(q_{ki}, q_{lj}) \quad (4)$$

$$0 \le dissim(q_i, q_j) \le p$$

Thus, the closer $sim(q_i, q_j)$ (respectively, $dissim(q_i, q_j)$) is to $p$, the more $q_i$ and $q_j$ are considered similar (respectively, dissimilar).

We also define similarity (respectively, dissimilarity) measures between two query sets and within a query set. These measures are defined by Formulas 5, 6, 7 and 8.

$$sim(C_a, C_b) = \sum_{q_k \in C_a, q_l \in C_b} \delta_{sim}(q_k, q_l) \quad (5)$$

$$0 \le sim(C_a, C_b) \le card(C_a) \times card(C_a) \times p$$

$$dissim(C_a, C_b) = \sum_{q_k \in C_a, q_l \in C_b} \delta_{dissim}(q_k, q_l) \quad (6)$$

$$0 \le dissim(C_a, C_b) \le card(C_a) \times card(C_a) \times p$$

$$sim(C_a) = \sum_{q_k \in C_a, q_l \in C_b, k<l} \delta_{sim}(q_k, q_l) \quad (7)$$

$$0 \le sim(C_a) \le \frac{card(C_a) \times card(C_a) \times p}{2}$$

$$dissim(C_a) = \sum_{q_k \in C_a, q_l \in C_b, k<l} \delta_{dissim}(q_k, q_l) \quad (8)$$

$$0 \le dissim(C_a) \le \frac{card(C_a) \times card(C_a) \times p}{2}$$

### 3.2.2 Clustering

Clustering consists in determining a so-called natural partition $P_{nat}$ composed of objects (here, queries) that reflects the internal structure of data. This partition must be such as its clusters are composed of objects with high similarity and objects from different clusters present a high dissimilarity.

Based on the previously defined functions, a clustering quality measure $Q(P_h)$ can be built, formula 9.

$$Q(P_h) = \sum_{a=1\ldots z, b=1\ldots z, a<b} (sim(C_a, C_b) + \sum_{a=1}^{z} disim(C_a)) \quad (9)$$

This measure permits to capture the natural aspect of a partition. Hence, $Q(P_h)$ measures simultaneously similarities between queries within the same cluster of partition $P_h$ and dissimilarities between queries within different clusters. Thus, we can define $Q(P_h)$ as an homogeneity function for the same class and an heterogeneity function for different classes. Therefore, the partitions presenting a high intra-cluster homogeneity and a high inter-cluster disparity have a weak value of $Q(P_h)$ and thereby appear as the most natural.

We have selected the Kerouac algorithm [15] for the clustering phase. This algorithm indeed bears several interesting properties:

1. its computational complexity is quite low (log linear regarding the number of queries and linear regarding the number of attributes);
2. it can deal with a high number of objects (queries);
3. it can deal with distributed data;
4. it allows to integrate constraints within the clustering process.

This last characteristic is particularly interesting, since it provides us with a way to integrate constraints concerning the view merging process.

### 3.3 Cost models

The number of candidate views is generally as high as the input workload is large. Thus, it is not feasible to materialize all the proposed views because of storage space constraints. To circumvent this limitation, we propose to use cost models allowing to keep only the most pertinent views.

Figure 5 shows the typical structure of an XML view. In our context, it is composed of *Cell* elements. Each *Cell* is itself composed of *dimension* elements that contain *Group by* attributes and *fact* elements corresponding to the aggregate results. We propose cost models that estimate the size and storage cost of a given XML view.

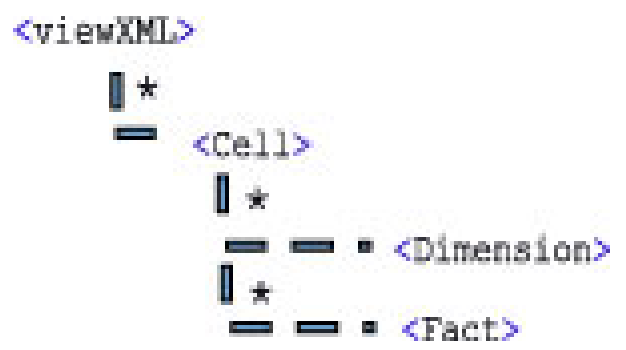

**Figure 5. XML view structure**

We estimate the size of a view by its number of elements. The number of Dimension and Fact elements in each Cell is the same. Indeed, the number of elements in a given view is estimated by the number of *Cell* elements. To compute it, we first estimate the maximum number of *Cell* elements (Formula 10).

$$ms(Cell) = \prod_{i=1}^{d} |d_i| \qquad (10)$$

$|d_i|$ is the cardinality of the dimension characterizing the *Cell* element. $d$ is the number of dimensions in the document *Dimensions.xml*.

Let $ms(v)$ be the maximum size of view $v$ that is composed of dimensions $d_1,...,d_k$, where $k$ is the number of dimensions in the view and $|d_i|$ the cardinality of dimension $d_i$. $ms(v)$ is expressed in Formula 11.

$$ms(v) = \prod_{i=1}^{k} |d_i| \qquad (11)$$

Golfarelli *et al.* [9] proposed to estimate the number of tuples in a given view *v* by using Yao's formula [26]. We also use this formula to estimate the number of *Cell* elements in *v* (Formula 12).

$$|v| = ms(v) \times \left[1 - \prod_{i=1}^{Cell} \frac{ms(Cell) \times c - i + 1}{ms(Cell) - i + 1}\right] \qquad (12)$$

$c = 1 - \frac{1}{ms(v)}$. If $\frac{ms(Cell)}{ms(v)}$ is large enough, this formula is well approximated by Cardena's formula [6]. Hence, we obtain Formula 13.

$$|v| = ms(v) \times \left(1 - \left(1 - \frac{1}{ms(v)}\right)^{Cell}\right) \qquad (13)$$

Cardenas and Yao's formulas are based on the assumption that data are uniformly distributed. The size, in bytes, of a view *v* is equal to the number of *Cell* elements multiplied by the average size needed to store one element. Thus, we estimate the size of a view as shown in Formula 14.

$$size(v) = |v| \times \sum_{i=1}^{k} size(d_i) \qquad (14)$$

$size(d_i)$ represents the size, in bytes, of dimension $d_i$ from *v* and *k* the number of dimensions.

### 3.4 Objective functions

We describe in this section three objective functions that help evaluating the variation of query execution cost induced by adding a new view. The query execution cost is assimilated to the number of *Cell* elements in the document *Facts.xml*, if no view is used; or to the number of *Cell* elements in the view(s) if they are exploited. The workload execution cost is obtained by adding the execution costs of each query within this workload.

The first objective function advantages the views providing more profit while executing queries. The second one advantages the views providing more benefit while occupying the smallest storage space. The third one combines the first two in order to first select all the views providing more profit and then retain only those occupying less storage space when this resource becomes critical. The first function is useful when the space storage is not limited, the second one is useful when storage space is small and the third one is interesting when storage is reasonably large.

#### 3.4.1 Profit objective function

Let $V = \{v_1,...,v_m\}$ be the candidate view set, $S$ the final view set and $Q = \{q_1,...,q_n\}$ a query set (workload). The profit objective function, noted *P*, is defined in Formula 15.

$$P_{/S}(v_j) = C_{/S}(Q) - C_{/S \cup vj}(Q) - \beta C_{update}(v_j) \qquad (15)$$

$$(v_j) \notin S$$

$C_{/S}(Q)$ denotes the query execution cost when all the views in $S$ are used. If this set is empty, $C_{/\emptyset}(Q) = |Q| \times |F|$ because all the queries are resolved by accessing fact $F$. When a view $v_i$ is added to $S$, $C_{/S \cup Vi}(Q) = \sum_{k=0}^{|Q|} C(q_k, v_j)$ denotes the query execution cost for the views that are in $S \cup v_i$. If query $q_k$ exploits $v_i$, cost $C(q_i, v_j)$ is then equal to $C_{vj}$ (number of tuples in $v_j$. Otherwise, $C(q_i, v_j)$ is equal to the maximum value between $F$ and value of $C(q_i, v)$ (executing cost of $q_i$ exploiting $v \in S$ with $v \neq v_j$).

Coefficient $\beta = |Q| p(v_i)$ estimates the number of updates for views $v_i$. The update probability $p(v_i)$ is equal to $\frac{1}{storage-space} \frac{\%update}{\%query}$, where the ratio $\frac{\%update}{\%query}$ represents the proportion of updating *vs.* querying the data warehouse.

$C_{update}(v_j)$ represents the maintenance cost for view $v_j$.

#### 3.4.2 Profit/space ratio objective function

If view selection is achieved under a space constraint, the profit/space objective function from Formula 16 is used. This function computes the profit provided by $v_j$ in regard to the storage space $size(v_j)$ it occupies.

$$R_{/S}(v_j) = \frac{P_{/S}(v_j)}{size(v_j)} \qquad (16)$$

#### 3.4.3 Hybrid objective function

The constraint on storage space may be relaxed if this space in relatively large. The hybrid objective function $H$ does not penalize space-greedy views if the ratio $\frac{remaining-space}{storage-space}$ is lower or equal than a storage-space given threshold $\alpha$, $0 < \alpha \leq 1$, where *remaining-space* and *storage-space* are respectively the remaining space after adding $v_i$ and the allotted space needed for storing all the views. This function is computed by combining the two functions $P$ and $R$ as shown in Formula 17.

$$H_{/S}(v_j) = \begin{cases} P_{/S}(v_j) \text{ if } \frac{remaining\text{-}space}{storage-space} \\ R_{/S}(v_j) \text{ otherwise} \end{cases} \qquad (17)$$

### 3.5 View selection algorithm

Our view selection algorithm (Algorithm 1) is based on a greedy search within the candidate view set $V$. The objective function $F$ must be one of the functions $P$, $R$ or $H$ described in the previous section. If $R$ is used, we add to the algorithm's input the storage space $M$ allotted for views. If $H$ is used, we also add threshold $\alpha$ as input.

**Algorithm 1** *View_Configuration_Construction*

```
S ← ∅
repeat
  v_max ← ∅
  F_max ← 0
  for all v_j ∈ V − S do
    if F_/S(v_j) > F_max then
      F_max ← F_/S(v_j)
      v_max ← v_j
    end if
  end for
  if F_/S(v_max) > 0 then
    S ← S ∪ {v_max}
  end if
until (F_/S(v_max) ≤ 0 or V − S = ∅)
```

In the first algorithm iteration, the values of the objective function are computed for each view within $V$. The view $v_{\max}$ that minimizes $F$, if it exists ($F_{/S}(v_{\max}) > 0$), is then added to $S$. If $R$ or $H$ is used, the whole storage space $M$ is decreased by the amount of space occupied by $v_{\max}$.

The function values of $F$ are then computed for each remaining view in $V - S$, since they depend on the selected views present in $S$. This helps taking into account the interactions that probably exist between the views.

We repeat these iterations until there is no improvement $(F_{/S}(v) \leq 0)$ or until all the views have been selected $(V - S = \varnothing)$. If functions $R$ or $H$ are used, the algorithm also stops when storage space is full.

## 4. Experiments

In order to validate our approach for XML materialized view selection, we generated an XML data warehouse, modeled according to the XCube specifications. This classical test data warehouse is composed of sales facts characterized by the *amount* and *quantity* measures. The facts are stored in the document *Facts.xml* (4,92 MB). They are described by five dimensions: *channels, promotions, customers, products* and *times* that are stored in the document *Dimensions.xml* (3.77 MB). This data warehouse has been implemented within the eXist native XML DBMS [17], which is a free tool that allows the storage of large documents and supports the XQuery language. We ran our tests on a Pentium 2 Ghz PC with 1 GB main memory and an IDE hard drive.

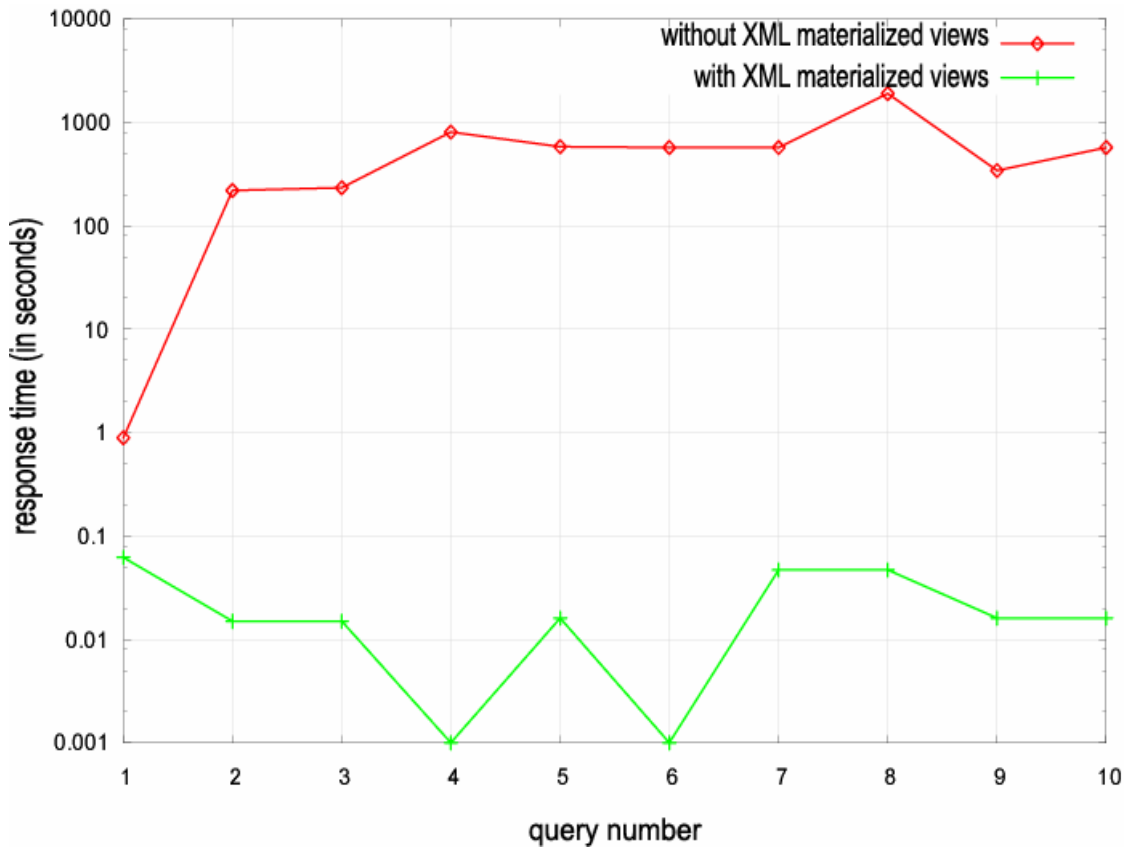

**Figure 7. Experimental results**

We executed on our data warehouse a workload composed of ten XQuery decision-support queries, with and without using our strategy. The selected views are stored in an independent collection. This collection is targeted by rewritten queries according to view data. We plotted in Figure 7 the execution time of our query workload on the original XML documents and on the materialized views we generated. The X-axis represents the ten queries and the Y-axis the corresponding execution time. The Y-axis is represented in logarithmic scale to highlight the difference between the execution costs. On an average, our XML view materializing strategy improves response time by a factor 24,700.

## 5. Conclusion and perspectives

In this paper, we have presented a strategy for materialized view selection in XML data warehouses. Our strategy exploits the results of clustering applied on a given workload to build a set of syntactically relevant candidate views. With the help of cost models we specifically designed for the XML model, we retain only the most advantageous candidate views. These models estimate data access cost using materialized views and storage cost for these views.

We have also proposed three objective functions: profit, profit/space ratio and hybrid that exploit our cost models to evaluate the execution cost of the workload. These functions are themselves exploited by a greedy algorithm that recommends a pertinent configuration of materialized views. This allows our strategy to respect the storage space constraint.

Finally, note that our strategy is independent from the warehouse model and the DBMS it is stored in. Though we used an XCube-based reference data warehouse, our strategy could easily be applied on any other model. In the same way, any DBMS could be used instead of eXist, including relational, XML compatible DBMSs.

The first experimental results we achieved are very encouraging, and show that our strategy guarantees a substantial gain in performance. However, our first perspective is to complement these results with other tests, possibly on other systems than eXist, and to assert in each configuration the gain in performance vs. the overhead for generating and refreshing the materialized views.

This work also opens two other axes of research perspectives. First, it is blatantly crucial to adapt or develop highly efficient optimization techniques in native XML DBMSs if they are to approach the performances of relational systems. XML indices are getting more and more efficient, but there is still room for improvement (*e.g.*, multidocument join indices). The generalized exploitation of materialized views could also be very beneficial. Thus, a rewriting query engine and refreshing strategies should be devised.

Our second research axis is even more specific to XML data warehouses. Decision-support queries bear specific needs in terms of operators. For instance, we had to extend XQuery to allow multiple *Group by* clauses to be able to implement our decision-oriented workload within eXist. Similar extensions do exist already, but it could be interesting to further extend XQuery to support OLAP operators such as *Cube, Rollup* or *Drill down*.